\documentclass[aps,prd,onecolumn,groupedaddress,showpacs,nofootinbib,amssymb
]{revtex4}
\usepackage[dvips]{graphicx}
\usepackage{amssymb}
\usepackage{amsmath}
\usepackage{graphicx}
\usepackage{amsfonts}
\usepackage{bm}

\begin{document}

\title{Inverse Symmetric Inflationary Attractors}
\author{
S.~D.~Odintsov,$^{1,2}$\,\thanks{odintsov@ieec.uab.es}
V.~K.~Oikonomou,$^{3,4}$\,\thanks{v.k.oikonomou1979@gmail.com}}
\affiliation{ $^{1)}$ICREA, Passeig Luis Companys, 23, 08010 Barcelona, Spain\\
08193 Cerdanyola del Valles, Barcelona, Spain\\
$^{2)}$ Institute of Space Sciences (IEEC-CSIC)\\
C. Can Magrans s/n, 08193 Barcelona, SPAIN\\
$^{3)}$ Laboratory for Theoretical Cosmology, Tomsk State University
of Control Systems and Radioelectronics (TUSUR), 634050 Tomsk,
Russia\\
$^{4)}$Tomsk State Pedagogical University, 634061 Tomsk, Russia \\
}

\begin{abstract}
We present a class of inflationary potentials which are invariant under a special symmetry, which depends on the parameters of the models. As we show, in certain limiting cases, the inverse symmetric potentials are qualitatively similar to the $\alpha$-attractors models, since the resulting observational indices are identical. However, there are some quantitative differences which we discuss in some detail. As we show, some inverse symmetric models always yield results compatible with observations, but this strongly depends on the asymptotic form of the potential at large $e$-folding numbers. In fact when the limiting functional form is identical to the one corresponding to the $\alpha$-attractors models, the compatibility with the observations is guaranteed. Also we find the relation of the inverse symmetric models with the Starobinsky model and we highlight the differences. In addition, an alternative inverse symmetric model is studied and as we show, not all the inverse symmetric models are viable. Moreover, we study the corresponding $F(R)$ gravity theory and we show that the Jordan frame theory belongs to the $R^2$ attractor class of models. Finally we discuss a non-minimally coupled theory and we show that the attractor behavior occurs in this case too.
\end{abstract}


\maketitle



\def\pp{{\, \mid \hskip -1.5mm =}}
\def\cL{\mathcal{L}}
\def\be{\begin{equation}}
\def\ee{\end{equation}}
\def\bea{\begin{eqnarray}}
\def\eea{\end{eqnarray}}
\def\tr{\mathrm{tr}\, }
\def\nn{\nonumber \\}
\def\e{\mathrm{e}}

\section{Introduction}

The inflationary paradigm is one of the most successful descriptions
of the early-time evolution of our Universe, in the context of
which, many shortcomings of the Big Bang cosmology are successfully
resolved \cite{inflation2,inflation3,inflation4,inflation5}. In a
recent study \cite{alpha1}, a new class of models was introduced and
these models are now known as $\alpha$-attractors models. These
models are based on a slow-rolling canonical scalar field with a
characteristic functional form of the inflationary potential. The
most appealing feature of this new class of cosmological models is
that the resulting spectral index of primordial curvature
perturbations and the scalar-to-tensor ratio is common to all the
models belonging to the general class of the $\alpha$-attractors
models, for large values of the $e$-foldings number $N$. Later
studies on these models
\cite{alpha2,alpha3,alpha4,alpha5,alpha6,alpha7,alpha8,alpha9,alpha10,alpha10a,alpha11},
and also some earlier studies in the same context
\cite{slowrollsergei}, indicated that several well-known canonical
scalar field models like the Starobinsky model
\cite{starob1,starob2}, and the Higgs model \cite{higgs}, are
limiting cases of the $\alpha$-attractors models.  In addition, the
fact that the latest Planck data \cite{planck} indicate that the
latter two models are in concordance with the observational
constraints, render the $\alpha$-attractors models quite timely and
intriguing models. Intriguing because there seems to be a common
origin behind all the viable cosmological models of inflation. In
fact, all the $\alpha$-attractors models potentials have a large
plateau, for large values of the canonical scalar field, and all
these models are asymptotically quite similar to the hybrid
inflationary scenario \cite{hybrid}. Another important feature of
the $\alpha$-attractors models is that in many cases these are
supergravity originating, and in most cases supersymmetry breaking
occurs at the minimum of the potential \cite{susybr1}. Interestingly
enough, the late-time acceleration era can find a successful
explanation in the context of $\alpha$-attractors models
\cite{linderefs1,linder}. Also for an interesting approach with
regards to a renormalization group Higgs potential, see
\cite{vernov}.

In this paper we shall present another interesting class of potentials which may lead to the same observational indices with the $\alpha$-attractors. This means that this new class of models, to which we refer to as ``inverse symmetry attractor models'', belongs to the $\alpha$-attractors models, if certain requirements are met. The inverse symmetry attractor models have a symmetry property related to a deformation parameter $\beta$ which appears in the potential. Particularly, the potentials are invariant under the transformation $\beta \to \frac{1}{\beta}$ and this actually justifies the name ``inverse symmetry attractors''. Special cases of this class of models, were originally studied in the context of holographic models \cite{holo1,holo2}, and also see Ref. \cite{hologr} for a recent study. In this paper we are interested in some special cases of these inverse symmetry attractors, which have the interesting property of belonging to the $\alpha$-attractors models. We shall investigate in detail when this property occurs and also we thoroughly discuss all the different cosmological inflationary scenarios that these models imply. In all cases we shall calculate the slow-roll indices and the corresponding observational indices and we compare the results to the latest observational data. As we will show, not all the inverse symmetry models lead to results compatible with the observations. However, it is intriguing that the $\alpha$-attractors related inverse symmetry models are, in most cases, compatible with the observational data. Finally, for the latter models, we shall find the corresponding $F(R)$ gravity equivalent theory and we thoroughly investigate whether the attractor property remains in the Jordan frame too. Finally we study a non-minimally coupled scalar theory which is symmetric under the transformation $\beta \to \frac{1}{\beta}$, and as we show, the attractor behavior occurs in this case too.

This paper is organized as follows: In section II we present the essential features of inverse symmetric attractors and we investigate under which conditions these models can belong to the $\alpha$-attractors class of models. To this end we calculate and study the slow-roll indices and the corresponding observational indices in various limits, and we critically discuss the results. In section III we compare the inverse symmetry attractors to the $\alpha$-attractors and we discuss the qualitative differences of the two classes of models. Also we present another inverse symmetry model and we study its cosmological phenomenological implications. In section IV we study the $F(R)$ gravity corresponding theory of the inverse symmetry attractor model we studied in section II and we investigate whether the attractor property occurs in the Jordan frame too. In section V we study a non-minimally coupled scalar-tensor theory and we demonstrate that the attractor behavior occurs in this case too. Finally, the conclusions follow in the end of the paper.

\section{Essential Features of Inverse Symmetric Attractors and Slow-roll Inflation}

In the following we shall consider a canonical scalar field theory, with the gravitational action being of the form,
\begin{equation}\label{canonocclact}
\mathcal{S}=\sqrt{-g}\left(\frac{R}{2}-\frac{1}{2}\partial_{\mu}\phi \partial^{\mu}\phi -V(\varphi))\right)\, .
\end{equation}
The geometric background we shall use in this paper is a flat Friedmann-Robertson-Walker metric, with line element,
\be
\label{metricfrw} ds^2 = - dt^2 + a(t)^2 \sum_{i=1,2,3}
\left(dx^i\right)^2\, ,
\ee
where $a(t)$ denotes the scale factor. Also we assume that the connection is a metric compatible and torsion-less, symmetric   affine connection, the Levi-Civita connection. Finally we use a physical units system such that $\hbar=c=8\pi G=\kappa^2=1$.

The focus in this paper is on inflationary potentials that depend on a free parameter $\beta$ and have the symmetry $\beta\to \frac{1}{\beta}$. One particularly interesting potential is the following:
\begin{equation}\label{mainpot}
V(\varphi)=\frac{\mu
(\beta+\frac{1}{\beta})}{1-\beta^m}\left(1-\beta^m-2ne^{-\beta\sqrt{\frac{2}{3}}\varphi}+2n\beta^me^{-\frac{1}{\beta}\sqrt{\frac{2}{3}}\varphi}\right)\,
,
\end{equation}
and in the following we shall extensively study the inflationary features of this potential. Also we shall discuss the similarities of this potential with the inflationary attractors described in Refs. \cite{alpha2,alpha3,alpha4,alpha5,alpha6,alpha7,alpha8,alpha9,alpha10}. The parameter $\mu$ in the potential (\ref{mainpot}) is an arbitrary mass scale, also $n$ is a freely chosen positive real number, while $\beta$ is a free parameter the values of which determine the behavior of the inflationary potential. Also the parameter $m$ is assumed to be a real positive number which satisfies $m\gg 1$. As it can be easily checked, the potential (\ref{mainpot}) is invariant under the transformation $\beta \to \frac{1}{\beta}$, and similar potentials were studied in Refs. \cite{holo1,holo2,hologr}. Actually, potentials with the symmetry $\beta \to \frac{1}{\beta}$ are related to holographic models, see Refs. \cite{holo1,holo2,hologr} and references therein. Let us now discuss the interesting features of the model (\ref{mainpot}), and firstly we discuss two limiting cases of the parameter $\beta$, namely the cases $\beta>1$ and $\beta<1$. Suppose that $\beta=\alpha$, and in the case $\beta=\alpha>1$, since $m\gg 1$, this implies that $\beta^m\gg 1$. In effect, for $\beta>1$ the potential becomes approximately equal to,
\begin{equation}\label{limit1}
V(\varphi)\simeq \mu \alpha \left(1-2ne^{-\frac{1}{\alpha}\sqrt{\frac{2}{3}}\varphi}\right)\, .
\end{equation}
This behavior is due to the fact that the term $e^{-\frac{1}{\alpha}\sqrt{\frac{3}{2}}\varphi}$ dominates over the term $e^{-\alpha\sqrt{\frac{3}{2}}\varphi}$, even for values of $\alpha$ a bit larger than one, for example $\alpha=4$ (we omit the physical units for simplicity). Notice that in this case $\alpha^m\gg 1$, so the limit we obtained is approximately correct. It would be helpful to see why the term $e^{-\frac{1}{\alpha}\sqrt{\frac{3}{2}}\varphi}$ dominates over $e^{-\alpha\sqrt{\frac{3}{2}}\varphi}$, for $\alpha>1$, so let us calculate these terms for $\alpha=11$ and for $\varphi=10^4$, where the canonical scalar field value should be chosen to be large, as it is expected during the inflationary phase, but the value we chose suffices to explain our argument. Then the aforementioned exponentials become,
\begin{equation}\label{expvalues}
e^{-\frac{1}{\alpha}\sqrt{\frac{2}{3}}\varphi}=5.8\times 10^{-11},\,\,\, e^{-\alpha\sqrt{\frac{2}{3}}\varphi}\simeq 2.5\times 10^{-3901}\, ,
\end{equation}
so this behavior clearly explains our argument. Now let us focus on the other limiting case, namely $\beta< 1$ which implies $\beta^m\ll 1$, so let us choose $\beta=\frac{1}{\alpha}$, with $\alpha$ satisfying $\alpha>1$, as in the previous case. In this case then, the potential becomes,
\begin{equation}\label{limit2}
V(\varphi)\simeq \mu \alpha \left(1-2ne^{-\frac{1}{\alpha}\sqrt{\frac{3}{2}}\varphi}\right)\, ,
\end{equation}
which is identical to the limiting case potential of Eq. (\ref{limit1}). Therefore, the two limiting cases, namely $\beta>1$ and $\beta <1$ yield the same potential, a result which to some extent was expected due to the symmetry $\beta\to \frac{1}{\beta}$. The result of the two limiting cases shows us that the limiting potential is identical to the attractor potentials of Refs. \cite{alpha2,alpha3,alpha4,alpha5,alpha6,alpha7,alpha8,alpha9,alpha10}, however there are some fundamental differences, as we show later on. Before we discuss this issue, let us firstly calculate the slow-roll indices and the corresponding observational indices for the limiting potential (\ref{limit1}).

\subsection{Slow-roll Era and Inflationary Indices}

Depending on the values of the parameter $\alpha$, the slow-roll era for the potential (\ref{limit1}) can yield quite different resulting expressions for the slow-roll indices and the corresponding observational indices. The slow-roll indices for a canonical scalar theory are equal to,
\begin{equation}\label{slowrollscalar}
\epsilon (\varphi)=\frac{1}{2}\left( \frac{V'(\varphi)}{V(\varphi)}\right)^2,\,\,\,\eta (\varphi)=\frac{V''(\varphi)}{V(\varphi)}\, ,
\end{equation}
and also the $e$-foldings number $N$ can be given in terms of the potential, during the slow-roll regime, so the resulting expression is,
\begin{equation}\label{efoldings}
N\simeq \int_{\varphi_f}^{\varphi_{i}}\frac{V(\varphi)}{V'(\varphi)}\mathrm{d}\varphi \, ,
\end{equation}
where the initial value of the canonical scalar field is $\varphi_i$, which we assume to be the value of $\varphi$ at the horizon crossing, and $\varphi_f$ is the value of the scalar field at the end of the inflationary era. The inflationary era ends when the first slow-roll index $\epsilon$ becomes of order $\mathcal{O}(1)$, so the condition $\epsilon\simeq 1$ yields,
\begin{equation}\label{slowrollend}
e^{\sqrt{\frac{2}{3\alpha^2}}\varphi_f}=2n+\sqrt{\frac{3\alpha^2}{4n^2}}\, .
\end{equation}
By integrating (\ref{efoldings}), for the potential (\ref{limit1}) we get the following expression at leading order,
\begin{equation}\label{leadingn}
N=-e^{\sqrt{\frac{2}{3\alpha^2}}\varphi_f}+e^{\sqrt{\frac{2}{3\alpha^2}}\varphi_i}\, ,
\end{equation}
so by substituting (\ref{slowrollend}) in (\ref{leadingn}) we get,
\begin{equation}\label{noneofthethree}
e^{\sqrt{\frac{2}{3\alpha^2}}\varphi_i}=2n+\frac{4nN}{3\alpha^2}+\frac{\sqrt{3}\alpha}{2n}\, .
\end{equation}
The slow-roll indices calculated at the horizon crossing $\varphi=\varphi_i$ are equal to,
\begin{equation}\label{slowhorexp}
\epsilon=\frac{4 n^2}{3 \alpha ^2 \left(e^{\frac{\sqrt{\frac{2}{3}} \varphi_i }{\alpha }}-2 n\right)^2},\,\,\,\eta=-\frac{4 n}{3 \alpha ^2 \left(e^{\frac{\sqrt{\frac{2}{3}} \varphi_i }{\alpha }}-2 n\right)}\, ,
\end{equation}
so by substituting Eq. (\ref{noneofthethree}) in Eq. (\ref{slowhorexp}), we have,
\begin{equation}\label{oposprepei}
\epsilon\simeq \frac{4 n^2}{3 \alpha ^2 \left(\frac{\sqrt{3} \alpha }{2 n}+\frac{4 n N}{3 \alpha ^2}\right)^2},\,\,\, \eta \simeq -\frac{4 n}{3 \alpha ^2 \left(\frac{\sqrt{3} \alpha }{2 n}+\frac{4 n N}{3 \alpha ^2}\right)}\, .
\end{equation}
We can easily calculate the observational indices of inflation, and particularly the spectral index of primordial curvature perturbations $n_s$ and the scalar-to-tensor ratio $r$, evaluated at the horizon crossing. The analytic expressions of $n_s$ and $r$ in terms of the slow-roll indices are given below,
\begin{equation}\label{spectscalindex}
n_s\simeq 1-6\epsilon+2\eta,\,\,\, r\simeq 16 \epsilon\, ,
\end{equation}
so by substituting (\ref{oposprepei}) in (\ref{spectscalindex}) we have,
\begin{equation}\label{generalobservindices}
n_s\simeq -\frac{8 n^2}{\alpha ^2 \left(\frac{\sqrt{3} \alpha }{2 n}+\frac{4 n N}{3 \alpha ^2}\right)^2}-\frac{8 n}{3 \alpha ^2 \left(\frac{\sqrt{3} \alpha }{2 n}+\frac{4 n N}{3 \alpha ^2}\right)}+1,\,\,\,r\simeq \frac{64 n^2}{3 \alpha ^2 \left(\frac{\sqrt{3} \alpha }{2 n}+\frac{4 n N}{3 \alpha ^2}\right)^2}\, .
\end{equation}
At this point, the resulting behavior of the inflationary observational indices strongly depends on whether $\alpha^2 >N$ or $\alpha^2<N$ and also on the values of $n$. We assume that $N\sim 50$ and also that $n>1$, so the case $\alpha^2>N$ can be realized for $\alpha>8$, in which case the observational indices are approximately equal to,
\begin{equation}\label{observindi}
n_s\simeq -\frac{32 n^4}{3 \alpha ^4}-\frac{16 n^2}{3 \sqrt{3} \alpha ^3}+1,\,\,\, r\simeq \frac{256 n^4}{9 \alpha ^4}\, .
\end{equation}
By choosing for example $(\alpha,n)=(35,8)$, the observational indices are,
\begin{equation}\label{observex1}
n_s\simeq 0.966289,\,\,\ r\simeq 0.0776399\, .
\end{equation}
The latest Planck data \cite{planck} indicate that the spectral index and the scalar-to-tensor ratio are constrained as follows,
\begin{equation}\label{observex1placj}
n_s\simeq 0.966,\,\,\ r< 0.10\, ,
\end{equation}
so the results (\ref{observex1}) are compatible with the Planck constraints (\ref{observex1placj}).

Let us now turn our focus on the case $N>\alpha^2$, and for $N=60$, the parameter $\alpha$ should be chosen to take values in the interval $(1,7)$. In this case, the observational indices of Eq. (\ref{generalobservindices}) are approximately equal to,
\begin{equation}\label{newfnsecondcase}
n_s\simeq 1-\frac{2}{N}-\frac{9 \alpha ^2}{2 N^2},\,\,\,r\simeq \frac{12 \alpha ^2}{N^2}\, .
\end{equation}
We need to note the functional resemblance of the resulting observational indices (\ref{newfnsecondcase}) to the results of the attractors of Refs. \cite{alpha2,alpha3,alpha4,alpha5,alpha6,alpha7,alpha8,alpha9,alpha10}. We shall discuss this issue in detail in the next section. The observational indices (\ref{newfnsecondcase}) can be compatible with the Planck results (\ref{observex1placj}), for example by choosing $(\alpha,N)=(4.1,60)$, we have,
\begin{equation}\label{observefinalres}
n_s\simeq 0.966667,\,\,\ r\simeq 0.0533\, ,
\end{equation}
which are in good agreement with (\ref{observex1placj}). It is important to note that the values $(\alpha,N)=(4.1,60)$ are not the only ones for which compatibility with the observational data can be achieved. For example by choosing $(\alpha,N)=(1.1,55)$, the observational indices take the following values,
\begin{equation}\label{observefinalresqww}
n_s\simeq 0.963636,\,\,\ r\simeq 0.0048\, ,
\end{equation}
which are in good agreement with the Planck data.

\subsection{Comparison with the Ordinary $\alpha$-attractors}

Let us now compare the results we obtained in the previous section, with the results of the attractor models of Refs. \cite{alpha2,alpha3,alpha4,alpha5,alpha6,alpha7,alpha8,alpha9,alpha10}. We consider two types of models, namely the T-models and the E-models, with the first models having the following potential,
\begin{equation}\label{tmodels}
V(\varphi)=\alpha \mu \tanh^2 (\frac{\varphi}{\sqrt{6\alpha}})\, ,
\end{equation}
where $\mu$ is a positive mass scale. In addition, the E-models have the following potential,
\begin{equation}\label{potentialemodels}
V(\varphi )=\alpha \mu^2 \left( 1-e^{-\sqrt{\frac{2}{3\alpha}}\varphi}\right)^{2 n}\, ,
\end{equation}
with the parameter $n$ being a positive number. In the case of the E-models, the potential in the small $\alpha$ limit reads,
\begin{equation}\label{smallalphaemodelpot}
V(\varphi )\simeq \alpha \mu^2 \left(1- 2 n e^{-\sqrt{\frac{2}{3\alpha}}\varphi}\right)\, ,
\end{equation}
and it can be checked that the small-$\alpha$ limit of the T-models is a subcase of the potential (\ref{smallalphaemodelpot}). By calculating the observational indices for the limiting potential (\ref{smallalphaemodelpot}), always in the small-$\alpha$ limit, we obtain at leading order,
\begin{equation}\label{observslowroli1}
n_s\simeq 1-\frac{2}{N}-\frac{9\alpha}{2N^2},\,\,\,r\simeq \frac{12\alpha}{N^2}\, .
\end{equation}
By comparing the results of Eqs. (\ref{newfnsecondcase}) and (\ref{observslowroli1}), we can see that these are identical. However, the difference is that the observational indices of Eq. (\ref{newfnsecondcase}) are obtained for the parameter $\alpha$ satisfying $\alpha>1$, while the ones of Eq. (\ref{observslowroli1}) are obtained for $\alpha\ll 1$. This is the main difference between the inverse symmetric attractors and the ordinary attractors of Refs. \cite{alpha2,alpha3,alpha4,alpha5,alpha6,alpha7,alpha8,alpha9,alpha10}. Another difference is that in the case of the ordinary attractors, for $\alpha=n=1$, the potential (\ref{potentialemodels}) becomes,
\begin{equation}\label{starobmodel}
V(\varphi )=\alpha \mu^2 \left( 1-e^{-\sqrt{\frac{2}{3}}\varphi}\right)^{2}\, ,
\end{equation}
which is the potential of the Starobinsky model \cite{starob1}. In the case of the inverse symmetric attractors, the limit $\beta=1$ cannot be taken, since divergences occur. In the next section we shall discuss this issue in some detail.

\section{Relation of Inverse Symmetric Potentials with the Exponential $R^2$ Model and Other Models}

Unlike the ordinary attractor models of Refs. \cite{alpha2,alpha3,alpha4,alpha5,alpha6,alpha7,alpha8,alpha9,alpha10}, the inverse symmetric potential of Eq. (\ref{mainpot}) cannot yield, at first sight at least, the Starobinsky model \cite{starob1,starob2}, since the limit $\beta \to 1$ makes the potential (\ref{mainpot}) singular. In this section we shall study the behavior of the potential (\ref{mainpot}), in the limit $\beta \to 1$. To this end we assume that $\beta=1+\varepsilon$, with the parameter $\varepsilon$ being an infinitely small number $\varepsilon \ll 1$. So in order to find the behavior of the potential near $\beta=1$, we will substitute $\beta=1+\varepsilon$ in the potential (\ref{mainpot}) and we expand in powers of $\varepsilon$ for $\varepsilon\to 0$. By doing so we obtain the following expansion,
\begin{align}\label{asxetonbod}
& V(\varphi)\simeq 2 \mu -\frac{8 \sqrt{\frac{2}{3}} \mu  n e^{-\sqrt{\frac{2}{3}} \varphi } \varphi }{m}-4 \mu  n e^{-\sqrt{\frac{2}{3}} \varphi }\\ \notag &
+\frac{\varepsilon ^2 \mu  e^{-\sqrt{\frac{2}{3}} \varphi } \left(-6 \sqrt{6} m^2 n \varphi -36 m n \varphi ^2+18 \sqrt{6} m n \varphi -54 m n+27 m e^{\sqrt{\frac{2}{3}} \varphi }-8 \sqrt{6} n \varphi ^3+72 n \varphi ^2-48 \sqrt{6} n \varphi \right)}{27 m}\, ,
\end{align}
where we omitted higher order terms in $\varepsilon$, since these are subdominant in the limit $\varepsilon\to 0$. By keeping the leading order term, the potential is,
\begin{align}\label{asxetonbod1}
V(\varphi)\simeq 2 \mu -\frac{8 \sqrt{\frac{2}{3}} \mu  n e^{-\sqrt{\frac{2}{3}} \varphi } \varphi }{m}-4 \mu  n e^{-\sqrt{\frac{2}{3}} \varphi }\, ,
\end{align}
and since $m\gg 1$, the last term can be omitted. Hence, in the limit $\varepsilon\to 0$, the potential can be approximated as follows,
\begin{equation}\label{seriespot}
 V(\varphi) \simeq 2 \mu  \left(1-2 n e^{-\sqrt{\frac{2}{3}} \varphi }\right)\, .
\end{equation}
The resulting approximate form of the potential (\ref{seriespot}), functionally looks like the Starobinsky model in the limit $\varphi \to \infty$, however it is different due to the factor ``$2$''. Hence, the functional behavior of the potential (\ref{mainpot}) in the limit $\beta \to 1$, looks like the Starobinsky model functionally, but it is not the same. This behavior of the inverse symmetry potential (\ref{mainpot}) can be regarded as another difference between the ordinary attractors and the inverse symmetry attractor potentials.

A similar model to the one appearing in Eq. (\ref{mainpot}) is the following canonical scalar field model, with its potential being,
\begin{equation}\label{alternpot}
V(\varphi)=\frac{2\mu}{3(1-\beta^m)}\left( e^{-\beta \sqrt{\frac{2}{3}}\varphi}-\beta^m e^{-\sqrt{\frac{2}{3}}\frac{1}{\beta}\varphi}\right)\, ,
\end{equation}
where $m\gg 1$ and the parameters $\mu$ and $\beta$ being chosen as the ones of the model (\ref{mainpot}). The model (\ref{alternpot}) is almost identical to the holographic-inspired model of Ref. \cite{hologr}, so in this section we shall investigate the phenomenological implications of this model. The model (\ref{alternpot}) is symmetric under the transformation $\beta\to \frac{1}{\beta}$, and in the case $\beta>1$, and by assuming that $\beta=\alpha$, with $\alpha>1$, the potential can be approximated as,
\begin{equation}\label{approx1}
V(\varphi)\simeq \frac{2\mu}{3}e^{-\frac{1}{\alpha}\sqrt{\frac{2}{3}}\varphi}\, ,
\end{equation}
and in the case $\beta<1$, where in this time $\beta=\frac{1}{\alpha}$, the potential is again approximated by the expression in Eq. (\ref{approx1}). So the two different limits of the parameter $\beta$ yield the same resulting limiting potential, and this is due to the fact that the potential is symmetric under $\beta\to \frac{1}{\beta}$. Let us now calculate the slow-roll indices for this limiting potential and also the corresponding observational indices. The slow-roll indices for the limiting potential (\ref{approx1}) have a particularly simple form,
\begin{equation}\label{symmetr}
\epsilon \simeq \frac{1}{3 \alpha ^2},\,\,\,\eta\simeq \frac{2}{3 \alpha ^2}\, ,
\end{equation}
and the corresponding observational indices are,
\begin{equation}\label{observfin}
n_s\simeq 1-\frac{2}{3 \alpha ^2},\,\,\, r\simeq \frac{16}{3 \alpha ^2}\, .
\end{equation}
It is obvious that the case at hand is much more constrained in comparison to the model (\ref{mainpot}), since the resulting observational indices depend solely on one parameter $\alpha$. For $\alpha=4.5$ the spectral index $n_s$ and the scalar-to-tensor ratio read,
\begin{equation}\label{spectral}
n_s\simeq 0.963944,\,\,\,r\simeq 0.288444\, ,
\end{equation}
so in this case it is not possible to obtain compatibility with the Planck data. However, for $\alpha\gg 1$, one obtains an exactly scale invariant spectrum with an extremely small scalar-to-tensor ratio. In Table \ref{table1} we have gathered the results for the models (\ref{mainpot}) and (\ref{alternpot}).
\begin{table*}[h]
\small
\caption{\label{table1}Two Inverse Symmetric Canonical Scalar Field Models and Their Phenomenology}
\begin{tabular}{@{}crrrrrrrrrrr@{}}
\tableline
\tableline
\tableline
Canonical Scalar Field Potential  & Limiting Potential for $\beta>1$ and $\beta<1$ & Observational Indices
\\\tableline
 $V(\varphi)=\frac{\mu (\beta+\frac{1}{\beta})}{1-\beta^m}\left(\beta^m-1-2ne^{-\beta\sqrt{\frac{2}{3}}\varphi}+2n\beta^me^{-\frac{1}{\beta}\sqrt{\frac{2}{3}}\varphi}\right)$ & $V(\varphi)\simeq \mu \alpha \left(1-2ne^{-\frac{1}{\alpha}\sqrt{\frac{3}{2}}\varphi}\right)\,\,\,\,\,\,$ & $n_s\simeq 1-\frac{2}{N}-\frac{9 \alpha ^2}{2 N^2},\,\,\,r\simeq \frac{12 \alpha ^2}{N^2}$
\\\tableline
$V(\varphi)=\frac{2\mu}{3(1-\beta^m)}\left( e^{-\beta \sqrt{\frac{2}{3}}\varphi}-\beta^m e^{-\sqrt{\frac{2}{3}}\frac{1}{\beta}\varphi}\right)$  & $V(\varphi)\simeq \frac{2\mu}{3}e^{-\frac{1}{\alpha}\sqrt{\frac{2}{3}}\varphi}$ & $n_s\simeq 1-\frac{2}{3 \alpha ^2},\,\,\, r\simeq \frac{16}{3 \alpha ^2}$
\\\tableline
\tableline
\tableline
 \end{tabular}
\end{table*}
As a conclusion, we can firmly say that not all the inverse symmetric models result to an attractor behavior, but it is certain that the models which in some limit are equal to the potential of Eq. (\ref{limit1}), then it is possible that these lead to the attractor observational indices (\ref{newfnsecondcase}).

\section{The $F(R)$ Gravity Description of Symmetric Attractors}

Having described the Einstein frame picture, let us now investigate the $F(R)$ gravity \cite{reviews1,reviews3,reviews4} equivalent theory corresponding to the potential (\ref{limit1}). Before going to the details of the calculation, let us recall in brief some essential features of the Einstein-Jordan frame correspondence, for details we refer to Refs. \cite{reviews1,slowrollsergei,sergeioikonomou1,sergeioikonomou2,sergeioikonomou4} and references therein. We consider the following $F(R)$ gravity action,
\begin{equation}\label{pure}
\mathcal{S}=\frac{1}{2}\int\mathrm{d}^4x \sqrt{-\hat{g}}F(R)\, ,
\end{equation}
with $\hat{g}_{\mu \nu}$ being the metric tensor in the Jordan frame. By introducing the auxiliary scalar field $A$ in the gravitational action (\ref{pure}), the action can be written as follows,
\begin{equation}\label{action1dse111}
\mathcal{S}=\frac{1}{2}\int \mathrm{d}^4x\sqrt{-\hat{g}}\left ( F'(A)(R-A)+F(A) \right )\, .
\end{equation}
Varying Eq. (\ref{action1dse111}) with respect to the auxiliary field $A$, we obtain the solution $A=R$, so in effect, the actions (\ref{action1dse111}) and (\ref{pure}) are mathematically equivalent. Then, in order to connect the Jordan to the Einstein frame, we perform the following canonical transformation,
\begin{equation}\label{can}
\varphi =\sqrt{\frac{3}{2}}\ln (F'(A))
\end{equation}
where $\varphi$ is the canonical scalar field in the Einstein frame. By making the following conformal transformation in the Jordan frame metric $\hat{g}_{\mu \nu }$,
\begin{equation}\label{conftransmetr}
g_{\mu \nu}=e^{-\varphi }\hat{g}_{\mu \nu }
\end{equation}
we can obtain the Einstein frame canonical scalar field action,
\begin{align}\label{einsteinframeaction}
& \mathcal{\tilde{S}}=\int \mathrm{d}^4x\sqrt{-g}\left ( R-\frac{1}{2}\left (\frac{F''(A)}{F'(A)}\right )^2g^{\mu \nu }\partial_{\mu }A\partial_{\nu }A -\left ( \frac{A}{F'(A)}-\frac{F(A)}{F'(A)^2}\right ) \right ) \\ \notag &
= \int \mathrm{d}^4x\sqrt{-g}\left ( R-\frac{1}{2}g^{\mu \nu }\partial_{\mu }\varphi\partial_{\nu }\varphi -V(\varphi )\right )
\end{align}
The detailed form of the canonical scalar field potential $V(\varphi)$ is equal to,
\begin{align}\label{potentialvsigma}
V(\varphi )=\frac{1}{2}\left(\frac{A}{F'(A)}-\frac{F(A)}{F'(A)^2}\right)=\frac{1}{2}\left ( e^{-\sqrt{2/3}\varphi }R\left (e^{\sqrt{2/3}\varphi} \right )- e^{-2\sqrt{2/3}\varphi }F\left [ R\left (e^{\sqrt{2/3}\varphi} \right ) \right ]\right )\, .
\end{align}
Then it is easy to find the $F(R)$ gravity which corresponds to the potential (\ref{potentialvsigma}), and this can be done by combining Eqs. (\ref{potentialvsigma}) and (\ref{can}). So by taking the derivative of Eq. (\ref{potentialvsigma}), with respect to the Ricci scalar, we obtain the following differential equation,
\begin{equation}\label{solvequation}
RF_R=2\sqrt{\frac{3}{2}}\frac{\mathrm{d}}{\mathrm{d}\varphi}\left(\frac{V(\varphi)}{e^{-2\left(\sqrt{2/3}\right)\varphi}}\right)
\end{equation}
with $F_R=\frac{\mathrm{d}F(R)}{\mathrm{d}R}$. By solving this with respect to $F_R$, yields the resulting $F(R)$ gravity which generates the potential $V(\varphi)$. By substituting the potential (\ref{limit1}) in Eq. (\ref{solvequation}), we obtain the following algebraic equation,
\begin{equation}\label{algegeneralalpha}
F_R =\frac{R}{4\alpha \mu}-F_R^{1-\frac{1}{\alpha}}(\frac{n}{\alpha}+2n)\, .
\end{equation}
This algebraic equation can be solved at leading order with respect to the Ricci scalar, and the parameter $\alpha$ plays a crucial role. Since $F_R\gg 1$ during the slow-roll inflationary era (this can be seen by looking at the canonical transformation (\ref{can})), we can see that since $\alpha>1$, the the exponent of the second term in the right hand side of Eq. (\ref{algegeneralalpha}), satisfies the following inequality $1-\frac{1}{\alpha}<1$. In effect, the second term in Eq. (\ref{algegeneralalpha}) is subleading, hence at leading order $F_R\simeq \frac{R}{4\alpha \mu}$. By substituting this in Eq. (\ref{algegeneralalpha}), we obtain that,
\begin{equation}\label{fr}
F(R)\simeq \frac{R^2}{8\alpha \mu}-\frac{\frac{n}{\alpha}+2n}{(2-\frac{1}{\alpha})(8\alpha \mu)^{}}R^{2-\frac{1}{\alpha}}+\Lambda\, ,
\end{equation}
where $\Lambda$ is an integration constant. In the following we shall assume that the $F(R)$ gravity at leading order is,
\begin{equation}\label{fror}
F(R)\simeq \frac{R^2}{8\alpha \mu}\, .
\end{equation}
Having the approximate expression for the $F(R)$ gravity, we can calculate the approximate expression for the Hubble rate in the slow-roll regime and eventually we can compute the observational indices. The equations of motion corresponding to the action  (\ref{pure}), for the FRW metric (\ref{metricfrw}), are equal to,
\begin{align}\label{cosmoeqns}
& 6F_RH^2=F_RR-F-6H\dot{F}_R ,\\ \notag &
-2\dot{H}F_R=\ddot{F}_R-H\dot{F}_R\, ,
\end{align}
Using the first differential equation in Eq. (\ref{cosmoeqns}), and by differentiating with respect to the cosmic time, we obtain the following differential equation,
\begin{equation}\label{oneofthefinals}
\frac{9 \gamma ^2 H(t) H^{(3)}(t)}{\mu }+\frac{27 \gamma ^2 H(t)^2 H''(t)}{\mu }+\frac{54 \gamma ^2 H(t) H'(t)^2}{\mu }=0\, ,
\end{equation}
and therefore by dividing with $H(t)^2$, we obtain,
\begin{equation}\label{enyaex}
\frac{9 \gamma ^2 H^{(3)}(t)}{\mu  H(t)}+\frac{27 \gamma ^2 H''(t)}{\mu }+\frac{54 \gamma ^2 H'(t)^2}{\mu  H(t)}=0\, .
\end{equation}
During the slow-roll era, the only dominant term in the above differential equation is the second one, so by solving the resulting equation we obtain the Hubble rate during the slow-roll era, which reads,
\begin{equation}\label{quasidesitter}
H(t)=H_0-H_i (t-t_k)\, ,
\end{equation}
which is a quasi-de Sitter evolution, as it was probably expected. Notice that $H_0$, $H_i$ in Eq. (\ref{quasidesitter}) are arbitrary integration constants, and $t_k$ is the time instance that the horizon crossing occurs during the inflationary era. Having the Hubble rate at hand, we can easily calculate the slow-roll indices, which in terms of the Hubble rate for a general $F(R)$ gravity are equal to \cite{noh},
\begin{equation}\label{slowrollparameters}
\epsilon_1=-\frac{\dot{H}}{H^2},\,\,\,\epsilon_2=0,\,\,\,\epsilon_3=\simeq \epsilon_1,\,\,\,\epsilon_4\simeq -3\epsilon_1+\frac{\dot{\epsilon}_1}{H(t)\epsilon_1}\, .
\end{equation}
In addition, the spectral index $n_s$ and the scalar-to-tensor ratio $r$ read,
\begin{equation}\label{nsdf}
n_s\simeq 1-6\epsilon_1-2\epsilon_4\simeq 1-\frac{2\dot{\epsilon}_1}{H(t)\epsilon_1},\,\,\,r=48\epsilon_1^2\, .
\end{equation}
Then by following the same description as in Ref. \cite{attractorsI}, we obtain the leading order $N$ observational indices, which are,
\begin{equation}\label{jordanframeattract}
n_s\simeq 1-\frac{2}{N},\,\,\,r\simeq \frac{12}{N^2}\, .
\end{equation}
The resulting picture is identical to the one obtained for the ordinary $R+\gamma R^2$ model at leading order. We need to note that the resulting observational indices are independent of $\alpha$.

\section{Non-minimal Coupling Description}

The approach we adopted for the Einstein frame theory can be extended in the case of a non-minimally coupled theory, in which case the gravitational action reads,
\begin{equation}\label{graviactionnonminimal}
\mathcal{S}=\int \mathrm{d}^4x\sqrt{-g}\left( \frac{f(\phi)R}{2\kappa^2}-\frac{1}{2}g^{\mu \nu}\partial_{\mu}\phi\partial_{\nu}\phi-V(\phi) \right)\, ,
\end{equation}
where the function $f(\phi)$ will be chosen to be symmetric under the transformation $\beta\to \frac{1}{\beta}$. The gravitational equations of motion for the action (\ref{graviactionnonminimal}), for a FRW metric, are equal to,
\begin{align}\label{gravieqnsnonminimal}
& \frac{3f}{\kappa^2}H^2=\frac{\dot{\phi}^2}{2}+V(\phi)-3h\frac{\dot{f}}{\kappa^2}\, ,\\ \notag &
-\frac{2f}{\kappa^2}\dot{H}=\dot{\phi}^2+\frac{\ddot{f}}{\kappa^2}-H\frac{\dot{f}}{\kappa^2}\, ,\\ \notag &
\ddot{\phi}+3H\dot{\phi}-\frac{1}{2\kappa^2}R\frac{\mathrm{d} f}{\mathrm{d}\phi}+\frac{\mathrm{d} V}{\mathrm{d}\phi}=0\, ,
\end{align}
where the ``dot'' denotes differentiation with respect to the cosmic time. For a non-minimally coupled scalar-tensor theory, the slow-roll indices are equal to,
\begin{equation}\label{slowrollnonminimal}
\epsilon_1=-\frac{\dot{H}}{H^2},\,\,\,\epsilon_2=\frac{\ddot{\phi}}{H\dot{\phi}},\,\,\,\epsilon_3=\frac{\dot{f}}{2Hf},\,\,\,\epsilon_4=\frac{\dot{E}}{2HE}\, ,
\end{equation}
where the function $E$ is in this case equal to,
\begin{equation}\label{epsilonparameter}
E=f+\frac{3\dot{f}^2}{2\kappa^2\dot{\phi}^2}\, .
\end{equation}
In terms of the slow-roll indices, and during the slow-roll era, the observational indices read,
\begin{equation}\label{observatinalindices1}
n_s\simeq 1-4\epsilon_1-2\epsilon_2+2\epsilon_3-2\epsilon_4,\,\,\,r=8\kappa^2\frac{Q_s}{f}\, ,
\end{equation}
where it is assumed that $\epsilon_i\ll 1$, $i=1,..,4$, and the parameter $Q_s$ is equal to,
\begin{equation}\label{qs1nonslowroll}
Q_s=\dot{\phi}^2\frac{E}{fH^2(1+\epsilon_3)^2}\, .
\end{equation}
We shall use the slow-roll conditions in order to find a simplified expression for the scalar-to-tensor ratio $r$, valid during the slow-roll era, so we will calculate $Q_s$ during the slow-roll era. The gravitational equations (\ref{gravieqnsnonminimal}) during the slow-roll era, are simplified as follows,
\begin{equation}\label{gravieqnsslowrollapprx1}
\frac{3fH^2}{\kappa^2}\simeq V(\phi)\, ,
\end{equation}
\begin{equation}\label{gravieqnsslowrollapprx2}
3H\dot{\phi}-\frac{6H^2}{\kappa^2}f'+V'\simeq 0\, ,
\end{equation}
\begin{equation}\label{gravieqnsslowrollapprx3}
\dot{\phi}^2\simeq \frac{H\dot{f}}{\kappa^2}-\frac{2f\dot{H}}{\kappa^2}\, ,
\end{equation}
where the ``prime'' denotes differentiation with respect to the scalar field $\phi$. The the parameter $Q_s$ takes the following form,
\begin{equation}\label{qs2}
Q_s=\frac{\dot{\phi}^2}{H^2}+\frac{3\dot{f}^2}{2\kappa^2fH^2}\, ,
\end{equation}
and by using the slow-roll equation (\ref{gravieqnsslowrollapprx3}), the parameter $Q_s$ is approximately equal to,
\begin{equation}\label{qs3}
Q_s\simeq \frac{H\dot{f}}{H^2\kappa^2}-\frac{2f\dot{H}}{\kappa^2H^2}\, .
\end{equation}
So by combining Eqs. (\ref{observatinalindices1}) and (\ref{qs3}), the scalar-to-tensor ratio finally reads,
\begin{equation}\label{rscalartensornoniminal}
r\simeq 16 (\epsilon_1+\epsilon_3)\, .
\end{equation}
Also in terms of the slow-roll indices, the spectral index of primordial curvature perturbations $n_s$ takes the following form,
\begin{equation}\label{nsintersmofslowroll}
n_s\simeq 1-2\epsilon_1\left (\frac{3H\dot{f}}{\dot{\phi}^2}+2 \right)-2\epsilon_2-6\epsilon_3\left( \frac{H\dot{f}}{\dot{\phi}^2}-1\right)\, .
\end{equation}
Having the above at hand we can proceed and specify the function $f(\phi)$ and we discuss the relation of the non-minimally coupled theory with the inverse symmetric attractors of the previous sections. A convenient choice for the function $f(\phi)$ is the following,
\begin{equation}\label{frphispec}
f(\phi)=\frac{1+\xi \left( e^{-\beta n\phi}+e^{-\frac{1}{\beta}n\phi}\right)}{2}\, ,
\end{equation}
where $\xi>0$ and $n>0$, while $\beta$ is chosen as in the previous sections. Also, the potential $V(\phi)$ can be chosen in such a way so  that it contains higher powers of the exponentials appearing in the function $f(\phi)$ in Eq. (\ref{frphispec}), but for the purposes of this paper, it suffices to choose the potential to be simply a constant, that is $V(\phi)=\Lambda$, with $\Lambda>0$. The function $f(\phi)$ in Eq. (\ref{frphispec}) is symmetric under the transformation $\beta \to \frac{1}{\beta}$, so by using the slow-roll formalism we developed in this paper, we shall calculate the slow-roll indices and the corresponding observational indices of this theory. We shall use a physical units system where $\kappa^2=1$ and also we assume that $\xi=1$ for simplicity, but the results are robust towards the choice of $\xi$. If $\beta>1$, then the function $f(\phi)$ is approximately equal to,
\begin{equation}\label{fphifirstmainapprox}
f(\phi)\simeq \frac{1+ e^{-\frac{n}{\beta}\phi}}{2}\, ,
\end{equation}
so by using the slow-roll approximated equations of motion, we have,
\begin{equation}\label{slowrollapproxeqnsofmotion}
\dot{\phi}\simeq -H\frac{n}{\beta}e^{-\frac{n}{\beta}\phi}\, .
\end{equation}
Also $\dot{f}$ in the slow-roll approximation reads,
\begin{equation}\label{slowrollapprxofdot}
\dot{f}\simeq \frac{n^2e^{-2\frac{n}{\beta}\phi }H}{2}\, .
\end{equation}
The first slow-roll index is,
\begin{equation}\label{firstdslowrollapproxanalyt}
\epsilon_1\simeq \left(\frac{\dot{\phi}}{H} \right)^2\frac{1}{2f}-\frac{\dot{f}}{2fH}\, ,
\end{equation}
so by using (\ref{slowrollapproxeqnsofmotion}) and (\ref{slowrollapprxofdot}), the first slow-roll index becomes,
\begin{equation}\label{epsilon1afteranalysis}
\epsilon_1\simeq \frac{n^2e^{-2\frac{n}{\beta}\phi }}{2\beta^2}\, .
\end{equation}
In the same way we can calculate the slow-roll indices $\epsilon_2$ and $\epsilon_3$, which are approximately equal to,
\begin{equation}\label{slowrolllast}
\epsilon_2\simeq \frac{n^2}{\beta^2}e^{-\frac{n}{\beta}\phi }+\epsilon_1,\,\,\,\epsilon_3\simeq \epsilon_1\, .
\end{equation}
So by combining Eqs. (\ref{rscalartensornoniminal}),  (\ref{nsintersmofslowroll}), (\ref{epsilon1afteranalysis}) and (\ref{slowrolllast}), we have,
\begin{equation}\label{nsandrprefinal}
n_s=1-2\frac{n^2}{\beta^2}e^{-\frac{n}{\beta}\phi },\,\,\,r\simeq 16 \frac{n^2}{\beta^2}e^{-2\frac{n}{\beta}\phi }\, .
\end{equation}
We can introduce the $e$-foldings number in the final expression for the observational indices (\ref{nsandrprefinal}), by using the following relation,
\begin{equation}\label{efolidngsanalyticresults}
N=\int_{t_k}^{t_f}H(t)\mathrm{d}t=\int_{\phi_k}^{\phi_f}\frac{H}{\dot{\phi}}\mathrm{d}\phi\simeq \frac{\beta^2}{n^2}e^{\frac{n}{\beta}\phi }\, ,
\end{equation}
where $t_k$ is the horizon crossing time instance and $\phi_k$ the value o the scalar field at that time, $t_f$ the time instance that inflation ends and $\phi_f$ the corresponding value of the scalar field. Also for the derivation of (\ref{efolidngsanalyticresults}) we used the fact that $\phi_k\gg \phi_f$, which is justified since during the slow-roll inflationary era, the values of the scalar field are quite large. From Eq. (\ref{efolidngsanalyticresults}) we have $\frac{n^2}{\beta^2}e^{-\frac{n}{\beta}\phi }=\frac{1}{N}$, so by substituting in Eq. (\ref{nsandrprefinal}), we have,
\begin{equation}\label{nsandrprefinal1}
n_s=1-\frac{2}{N},\,\,\,r\simeq  \frac{16\beta^2}{n^2N^2}\, .
\end{equation}
Finally, for $n=\frac{2}{\sqrt{3}}$, the observational indices read,
\begin{equation}\label{nsandrprefinal2}
n_s=1-\frac{2}{N},\,\,\,r\simeq  \frac{12\beta^2}{N^2}\, ,
\end{equation}
which are identical to the resulting observational indices of Eq. (\ref{newfnsecondcase}). Hence as we demonstrated, it is possible to have the attractor behavior even with a suitably chosen non-minimally coupled theory. Finally, we need to note that the non-minimally coupled theory we chose is different from the strongly coupled non-minimally coupled theory of Ref. \cite{lindelast}, where the function $f(\phi)$ has to satisfy a different class of criteria, and also $\xi$ should be quite large.

\section{Conclusions}

In this paper we studied some inflationary potentials with a special inverse symmetry. As we demonstrated, it is possible that these inverse symmetry models can belong to the more general class of $\alpha$-attractors models. As we showed, this is possible in some limits of the theory, however, there are some quantitative differences with the $\alpha$-attractors models. We investigated various limiting cases of the inverse symmetry models and we showed that in all cases, compatibility with the observational data is achieved. We discussed in some detail all the qualitative features of the inverse symmetric models of inflation and also we investigated whether these models can be limiting cases of the Starobinsky model, with the answer lying in the negative. Also we showed that not all inverse symmetric potentials yield quantitatively viable results, and as we showed, it seems that when the limiting inverse symmetric potential is identical to the one corresponding to the $\alpha$-attractor models, then compatibility with the observations can be achieved. Moreover, we found the $F(R)$ gravity theory corresponding to the inverse symmetry model and we calculated the observational indices in order to see if the attractor property remains. As we showed, the resulting indices are identical to the ones corresponding to the $R+\gamma R^2$ model, so the $R^2$ model is the attractor of this kind of potentials in the Jordan frame. Finally, we showed how the attractor behavior can occur in the case of a non-minimally coupled scalar-tensor theory.

What we did not address in this paper is the relation of the inverse symmetric attractors with supergravity superpotentials. The motivation for this study is that some of the inverse symmetric models are directly related to supergravity superpotentials \cite{holo1}, so what now remains is to investigate whether the $\alpha$-attractor related inverse symmetric potentials can be derived by supergravity. We defer this task to a future work.

\section*{Acknowledgments}

This work is supported by MINECO (Spain), project
 FIS2013-44881 (S.D.O) and by Min. of Education and Science of Russia (S.D.O
and V.K.O).

\end{document}